# Network structure and fragmentation of the Argentinean interbank markets

**Pedro Elosegui[1]**
Banco Central de la República Argentina

**Federico D. Forte**
BBVA Research, Banco BBVA Argentina

**Gabriel Montes-Rojas**
IIEP-BAIRES-UBA & CONICET

*Abstract*

This paper studies the network structure and fragmentation of the Argentinean interbank market. Both the unsecured (CALL) and the secured (REPO) markets are examined, applying complex network analysis. Results indicate that, although the secured market has less participants, its nodes are more densely connected than in the unsecured market. The interrelationships in the unsecured market are less stable, making its structure more volatile and vulnerable to negative shocks. The analysis identifies two "hidden" underlying sub-networks within the REPO market: one based on the transactions collateralized by Treasury bonds (REPO-T) and other based on the operations collateralized by Central Bank (CB) securities (REPO-CB). The changes in monetary policy stance and monetary conditions seem to have a substantially smaller impact in the former than in the latter "sub-market". The connectivity levels within the REPO-T market and its structure remain relatively unaffected by the (in some period pronounced) swings in the other segment of the market. Hence, the REPO market shows signs of fragmentation in its inner structure, according to the type of collateral asset involved in the transactions, so the average REPO interest rate reflects the interplay between these two partially fragmented sub-markets. This mixed structure of the REPO market entails one of the main sources of differentiation with respect to the CALL market.

---

[1] The authors appreciate the valuable comments of anonymous reviewers who have contributed to improve the original manuscript with their suggestions. The *Gerencia Principal de Estadísticas Monetarias* of BCRA and the *Gerencia de Mercado* of MAE, provided the information and valuable technical support. The information was anonymized and protected under strict confidentiality. The opinions expressed in this work are the responsibility of the authors and do not necessarily reflect those of the BCRA or its authorities.

## 1. Introduction

Interbank markets play a central role in financial systems. They are key for the monetary policy implementation by enhancing liquidity management operations of central banks and financial entities. Complementarily, the interbank interest rates, which are in general short-term rates (usually, overnight), act as major reference or benchmark for the rest of the interest rates in the economy. Hence, the "price signals" that emerge from these markets constitute a leading indicator of the prevailing monetary conditions in the system.

A significant distinction between different types of interbank markets lies on whether the operations are secured or unsecured, that is, if a collateral asset backs the transactions or not. If banks exchange liquidity on a secured basis, it means that other assets (for example, government bonds) are granted by the debtor entity to guarantee the loan. In contrast, in unsecured markets, interbank loans are not backed by any collateral, so the risk involved in the operations is higher. This crucial difference has implications in terms of the interest rates and maturities of the transactions, as well as on the overall functioning of the market itself. In addition, the secured interbank operations backed with government bonds and/or central bank securities provide market liquidity and price reference for those instruments, contributing to the depth and development of the domestic bond market.

Despite their centrality and importance in the financial system, interbank markets are sometimes "*fragmented*". In the case of well-functioning and efficient interbank markets, from an individual bank's risk perspective, there should be no difference between segments of the same market. However, the evidence shows that banks may face (risk-adjusted) differential funding costs or entry barriers in separate segments of interbank markets, which can have significant welfare costs (Gabrieli and Labonne, 2018a,b). This fragmentation hinders the smooth transmission of monetary policy and thus impair liquidity management and credit supply by affecting the funding capacity of banks and the price signals embodied by interbank rates. The absence of frictions between different segments of the markets are essential to reduce asset prices' volatility and stabilize the economy, hence the relevance of properly detecting potential fragmentation *across and within* interbank markets, which, in case of existence, should be in turn addressed by policy-makers and financial supervisors.

The aim of this study is to assess the actual fragmentation of the interbank markets in Argentina, as well as potential implications for monetary policy and financial stability. For this purpose, we go beyond the heterogeneity in interest rates faced by individual banks and apply network analysis methods to examine the underlying topological structure of the Argentinean markets. This approach allows the detection of structural differences between or within the markets, divergent dynamics of key topological measures, or dissimilar reactions across segments of the market when facing specific contexts or events, which could entail evidence of market fragmentation.

Hence, in this paper the Argentinean interbank markets are interpreted as networks, where the nodes are the active financial entities, while the links are the loans among them. There are two main interbank markets in the country. The secured market is called REPO market, while the unsecured market is known as CALL market. The Central Bank of Argentina (Banco Central de la República Argentina, BCRA) intervenes only in the REPO market. Both CALL and REPO interest rates represent important benchmark rates in the domestic financial system. Previous studies described their main basic characteristics. For instance, Anastasi, Blanco, Elosegui and Sangiácomo (2010) analyzed the impact of interbank relationships on access to liquidity, focusing on CALL market; Forte (2020) analyzed the fundamental aspects of the unsecured market's network topology; and Elosegui and Montes-Rojas (2020) studied the effect of local and global network measures on the interest rate spreads, finding heterogeneous effects in both the secured and unsecured markets. Based on a novel comparative analysis of the topological structures of both the secured and unsecured market (and their co-evolution over time), our main contribution in this paper is to provide evidence on the existence of fragmentation *between* them and/or *within each* market, resorting to graph theory and network analysis techniques.

The interbank markets can be understood as complex networks that connect numerous banks and other financial institutions through different types of exposures.[2] Also, these interactions take place in secured and unsecured interbank markets which have different institutional frameworks and procedures. Moreover, they also differ in market entry conditions and may be affected by particular market and banking regulations, if not different taxation treatments.

In our approach we make use of the relevant literature on network fragility. The distribution of the interconnections across the agents in a network has meaningful implications in terms of its stability and potential systemic risks. For instance, a financial network structure characterized by a high concentration of connections among a few banks is subject to significantly different risks than a network where the interconnections are more evenly distributed across participants.

Therefore, we analyze the empirical network degree distribution corresponding to the different segments of the Argentinean interbank market from 2015 to 2018, which allow us to draw conclusions on their comparative structure and evolution through different contexts. For analytical purposes, we not only distinguish the CALL market from the REPO market, but we also examine thoroughly the inner structure of the latter: we examine how its network structure changes when the Central Bank (CB) operations are excluded, and, additionally, how it changes when different types of collateral assets are involved (Central Bank securities or Treasury bonds). The approach provides a new perspective on how Argentinean banks interact in each segment and how regulations and market structures affect such behavior, with potentially relevant implications for aggregate interest rates in the markets and prudential regulation, liquidity management as well as financial market development. The analysis underscores the presence of two "hidden" underlying sub-networks within the REPO market according to the type of collateral involved in the operations. Both sub-markets react differently to equal contexts or monetary policy interventions, with low co-movement levels in some volatile periods. Changes in the monetary policy stance and monetary conditions seem to have a substantially different impact in each sub-market. This mixed structure of the REPO market is one of the main drivers of the intermittent uncorrelation with respect to the CALL market topological measures and interest rates.

---

[2] Prudential regulations emphasize the importance of monitoring the degree of interconnectedness among banks, in order to prevent potential macrofinancial risks and financial instability (BCBS, 2018).

For a better understanding the potential fragmentation, the network fragility in the different segments of the interbank market is assessed by studying the underlying characteristics of their network structures. We approximate the theoretical distribution that better fits the empirical degree distribution for both the unsecured and the secured market. With that objective, we apply the methodology developed by Clauset et al. (2009). This paper extends previous results in Forte (2020) that studied the degree distribution of the CALL market, which proved to be more compatible with a Lognormal than with Poisson or Power Law distributions[3]. This aspect tends to be particularly relevant in markets where interbank liquidity is traded predominantly in unsecured markets or in secured markets but with potentially illiquid and/or risky underlying assets. Moreover, potential liquidity shocks or a fall in the market value of a collateral are not unusual in these types of networks. Also, the access and interaction of banks with markets having different degree distributions may be reflecting their adaptive behavior to the segmented interbank market operation.

The paper is organized as follows. Section 2 briefly reviews the literature on interbank networks and the theory behind the analysis of the degree distribution as a proxy of network fragility. Section 3 describes the institutional framework of the Argentinean interbank markets, considering the different segments and their characteristics. Section 4 presents the empirical network analysis of the unsecured and secured markets and outlines the main results. Finally, section 5 discusses and summarizes the conclusions.

## 2. Network topology and fragility of interbank markets

There is a large literature that applies network analysis to study interbank markets, and more recently extended to capture the systemic exposure of financial institutions. As mentioned before, the interbank network plays a key role in the financial system and, from a macroprudential perspective, understanding its topological structure is relevant for both the Central Bank and the financial supervisor to determine its robustness and/or fragility to shocks. Complementarily the relative fragility of a network can be analyzed through the characterization of the underlying degree distribution.

Network analysis of the degree of interconnectedness in the financial system can inform policymakers on how regulation can prevent and/or reduce banking instability as well as on optimal bank resolutions mechanisms. Hence, empirical networks have been used for stress test exercises[4]. Network centrality measures, developed to assess centrality in other contexts and markets and adapted to the context of financial networks, can guide policy makers in their evaluation of the systemic importance of financial and non-financial institutions.

A number of papers investigate the interplay between financial distress and topological characteristics of interbank networks, focusing on the network resilience to different kinds of shocks.[5] In the case of Argentina, Forte (2020) analyses the network structure of the unsecured CALL market. The author found a short average distance between nodes and concluded that the market structure cannot be characterized as a random network. In this market, centrality variables lead to results compatible with the presence of embedded relationships among banks.[6] While some authors argue that a more interconnected architecture enhances the resilience of the system to the failure of an individual bank because credit risk is shared among more creditors, others suggest that a higher density of connections may function as a destabilizing force, facilitating financial distress to spread through the banking system. The overall picture that emerges from this literature is that the density of linkages has a non-monotonic impact on systemic stability and its effect varies with the nature of the shock, the heterogeneity of the players and the state of the economy. Thus, no optimal network structure that is more resilient under all circumstances can be identified.[7]

---

[3] Random networks' degree distributions tend to follow a Poisson (or exponential), while scale-free networks' degree distributions are best fitted by a Power Law (Barabási and Albert, 1999, Albert and Barabási, 2002).

[4] See Upper (2011) for a comprehensive review.

Network positioning could affect banks' interest rates through various mechanisms. First, in line with Acemoglu, Ozdaglar and Tahbaz-Salehi (2015), dense interconnections serve as a mechanism for the propagation of shocks, leading to a more fragile financial system. As such, banks that are more connected may be perceived by the market as fragile. The same banks can be perceived as *too-interconnected-to-fail* such that rather than fragile, those banks are perceived as more likely to be bailout.[8] This is similar to the *too-big-to-fail* effect observed in other interbank markets. Second, as argued by Booth, Gurun and Zhang (2014), financial institutions with more extensive and strategic financial networks, can more efficiently acquire and process information due to their better access to order flows. Third, banks with higher centrality within the network have better access to liquidity and are able to charge larger intermediation spreads; see for instance, Temizsoy, Iori, and Montes-Rojas (2015, 2017).[9]

The structure of interbank networks has been mapped for several countries, where the topology of interbank markets has been characterized and stylized facts and regularities have been identified.[10] The most common findings reported in this literature are: (i) interbank networks are sparse; (ii) degree and transaction volume distributions are fat-tailed, revealing heterogeneous players characteristics; (iii) the networks show disassortative mixing with respect to the bank size, so small banks tend to trade with large banks and vice versa; (iv) clustering coefficients are usually quite small; (v) interbank networks satisfy the small-world property;[11] (vi) interbank networks have a tiering structure with a tightly connected core of money-center banks to which all other periphery banks connect.

---

[5] See Iori, Jafarey and Padilla (2006); Nier, Yang, Yorulmazer and Alentorn (2007); Gai, Haldane and Kapadia (2011); Battiston, Puliga, Kaushik, Tasca and Caldarelli (2012); Karik, Gai and Marsili (2012); Lenzu and Tedeschi (2012); Georg (2013); Roukny, Bersini, Pirotte, Caldarelli, and Battiston (2013); Acemoglu, Ozdaglar and Tahbaz-Salehi (2015).

[6] Indeed, in a previous work, Anastasi, Blanco, Elosegui and Sangiácomo (2010) reported similar empirical results for the same market.

[7] For a survey on systemic risk and financial contagion see Chinazzi and Fagiolo (2013).

[8] See for instance Battiston, Puliga, Kaushik, Tasca and Caldarelli (2012).

[9] Previous empirical evidence as Angelini, Nobili and Picillo (2011), Bech, Chapman and Garratt (2010), Temizsoy, Iori and Montes-Rojas (2017) suggest that being systemically more important, in term of size or connectedness, can explain part of the cross-sectional variation in banks' borrowing costs before and during the global financial crisis.

[10] Examples include Boss et al. (2004) for the Austrian interbank market, Soramaki et al. (2007) and Bech and Atalay (2010) for the US Federal funds market, De Masi et al. (2006), Iori et al. (2008) and Fricke and Lux (2015) for the Italian-based e-MID, Degryse and Nguyen (2007) for Belgium, Craig and Von Peter (2014) for the German interbank market, Langfield et al. (2014) for the UK and in Veld and van Lelyveld (2014) for the Dutch market. Billio et al. (2012) studies the time-series properties of interconnectedness measures in financial markets.

In the specific case of Latin American countries, the interbank networks of Brazil (Silva, Souza and Tabak, 2016; Souza et al, 2014; Tabak et al, 2014; Cajueiro and Tabak, 2008) and Mexico[12] are the most extensively analyzed, but also recent papers have studied the interbank markets of Bolivia (Caceres-Santos, Rodriguez-Martinez, Caccioli and Martinez-Jaramillo, 2020), Colombia (León and Renneboog, 2014; León and Miguelez, 2021) and Perú (Cuba *et al.*, 2021). There are some contrasts between these different networks of the region, mainly related to specificities of each banking system, like foreign currency exposures and/or their size/depth. However, all of them share relevant network characteristics: a core-periphery structure, disassortative mixing, short average distances and clustering coefficients which entail useful indicators of systemic risks. Nevertheless, the concept of fragmentation has not been addressed yet with the approach implemented in this paper, which in fact could be easily applied to other Latin American or emerging countries to assess this phenomenon in their respective interbank markets.

The study of the markets' network structure provides some key insights about their stability when facing different kind of shocks. This issue leads us to another important literature strand related to the empirical degree distribution of the network, which is one of the most important elements that define the underlying topological structure of the system.

In random networks, the nodes' degree distribution tends to behave similarly to a Poisson (or exponential) distribution, while scale-free networks are better described by a Power Law (Albert and Barabási, 2002). A fat-tailed degree distribution (like the Power Law or the lognormal) implies that in such network a few highly connected nodes coexist with a myriad of low-connected agents. This fact has strong implications in terms of the resilience of the system, as those networks can be characterized as *robust-yet-fragile* structures (Albert and Barabási, 2000): they are surprisingly resilient against random errors, that is, to random failures or removals of a large number of nodes (robustness), even when the networks are faced with high failure rates, but this error tolerance is coupled with a high weakness to targeted attacks to the most central nodes of the network, as it rapidly breaks into isolated fragments when a few of the most connected nodes are removed (vulnerability). In contrast, networks that do not have such strong dominant central nodes tend to be more resistant to targeted shocks. In fact, random graphs present this converse risk structure. They easily absorb targeted attacks but tend to fall apart rapidly with random failures. This happens because those networks do not have particularly central nodes of systemic relevance that provide cohesion to the network structure.

---

[11] A network is *small-world* if the mean geodesic distance between pairs of nodes is small relative to the total number of nodes in the network, that is, this distance grows no faster than logarithmically as the number of nodes tends to infinity.

[12] Martinez-Jaramillo *et al.* (2014) characterized the Mexican interbank network; Molina-Borboa *et al.* (2015) and Poledna *et al.* (2015) studied the multi-layer network of exposures among Mexican banks, including interbank credit, securities, foreign exchange and derivative markets; while Usi-Lopez, Martinez-Jaramillo and López-Gallo (2017) analyzed the repo market in that country.

Hence, in the context of interbank networks, this attribute has key implications regarding the assessment of systemic fragility and potential risks. If an empirical financial network displays a fat-tailed behavior, then a rigorous identification of the central agents in the graph should become a priority task for central banks and regulators.

In terms of market fragmentation, it is interesting to note that Gabrieli and Labonne (2018a) showed for the case of European banks in the 2011-2015 period that the fragmentation in the interbank market was mainly explained by two sources: bank idiosyncratic risk and sovereign-dependence risk. In 2011, the ECB announced interventions with open market operations in secondary government bonds markets, the so called Outright Monetary Transactions (OMT) on sovereign debt securities. As a result of the increased secondary market liquidity and the ECB implicit collateral, the fragmentation in the interbank market declined. In fact, both sovereign and idiosyncratic risks identified by the authors were reduced by these two factors.

In this paper, we apply a statistical procedure developed by Clauset et al. (2009) to assess which is the theoretical distribution that best fits the empirical degree distribution of each market network. The authors consider the relative relevance of different fat-tailed distributions for the degree distribution at hand. For instance, the most crucial consequences of the Power Law or the Lognormal distribution of degrees derive from the fact that they are fat-tailed, in comparison with Poisson or Exponential distributions. Taking that issue into account, the authors test other types of fat-tailed distributions in addition to Power Laws, not limiting themselves just to that latter alternative. These distributions show histograms with a slower decay compared to an exponential distribution as the variable of interest increases (in our case, the node degree):

$$(1) \qquad \lim_{x \to \infty} \frac{f(x)}{e^{-x}} \neq 0$$

Therefore, in order to assess the network fragility in the Argentinean interbank market, we identify the theoretical distribution that better fit to the empirical degree distribution of both the secured and the unsecured market. In addition, we examine potential differences *within* the REPO market, including or excluding the BCRA from the network. Finally, we investigate if structural differences arise when considering the operations using Treasury bonds as collateral or those backed by BCRA securities. Results showing significantly different empirical distributions in the markets may be not only an indication of different risks from the macroprudential point of view, but also an evidence of market fragmentation in the interbank market.

### 3. The Argentinean interbank markets

During the period under analysis, the Argentinean banking system was comprised of 76 banks.[13] Five of them concentrated 50% of total credit and most of them actively participated in the interbank market (71 in the unsecured and 52 in the secured segment). The interbank market can be understood as a complex network where the interaction between banks and the BCRA[14] determines relevant reference interest rates for the economy: the unsecured or CALL rate, the secured or REPO rate, in deep interaction with the Central Bank's reference monetary policy interest rate.[15] The monetary policy is transmitted through the interbank interest rate to the rest of the interest rates of the financial system (deposit rates, loan rates and others) impacting on economic activity level and/or inflation.

---

[13] Including 12 public banks (2 national, 9 provincial and 1 municipal banks), 33 domestic banks, 9 foreign banks, 7 branches of foreign banks and 15 non-bank financial entities.

[14] The participant banks use the interbank market to negotiate temporary reserve positions (surplus or deficit), as well as securities (in secured markets), and to manage liquidity that may eventually be channeled to the non-financial sector. Also, the BCRA's monetary (and FX) operations, carried out for the fulfillment

The unsecured or CALL market is an over-the-counter market where banks can settle operations bilaterally. The participants have counterparty exposure limits that frame the bilateral transactions, subject to the general limits imposed by Central Bank regulations. The transactions are cleared on the *Medio Electrónico de Pagos* (MEP) platform of BCRA.[16] Most of the activity is concentrated in overnight loans, with only a small number of transactions maturing beyond three days (the weekends or extended holidays). Most of the banks, including the smaller and specialized ones, operate in this market. In fact, some of them only have access to the CALL market and do not operate in the REPO market. As mentioned before, the CB does not operate in this unsecured segment of the market. The overnight CALL rate is published by BCRA and it is a traditional benchmark rate for the financial market.

On the other hand, the secured or collateralized market, called REPO market, functions through the electronic trading platform SIOPEL of the *Mercado Abierto Electrónico* (MAE).[17] Only banks and financial entities that are members (or adherents) to MAE can participate in this platform, which explains why there are significantly less participant is this market compared to the CALL market. The platform is anonymous and bilateral with all positions visible by the participants. It is an order-driven system with no market-making arrangements. The transactions are not settled through a central clearing counterparty and each participant establishes counterparty limits, although the credit risk is limited by the use of collateral (treasury and/or central bank securities) and haircuts.[18] In fact, the system is settlement-risk free, since there is an online validation of the portfolio limits for each transaction between the parties.

The BCRA actively participates in the secured market through its lending and deposit facilities called "*Pases*" (the active transactions -loans to banks- are known as "REPOs", while the deposit facilities used by financial entities are called "Reverse REPOs"). Additionally, the Central Bank conducts open market operations (non-systematic and sparse). The BCRA issues its own debt securities, to absorb or provide liquidity from/to

---

of its objectives (including bonds operations, passive and active repos, open market operations, loans to financial institutions, and others), are implemented through debits and/or credits in the current accounts of the banks affecting the same reserves positions and their minimum reserves requirements compliance.

[15] The cut-off interest rate of the primary issuances of the BCRA's securities, together with the Central Bank's REPOs rates (deposit and credit facilities) are the reference or monetary policy rates in Argentina.

[16] The MEP is a Real Time Gross Settlement (RTGS) platform. Also, the net transactions are daily informed through SISCEN Information Task Requirement of the bank regulator (*Superintendencia de Entidades Financieras y Cambiarias*). The information of the CALL market used in this study comes from that source.

[17] The MAE is an electronic negotiation market created in 1989. It is the main electronic market for the negotiation of securities, foreign currency and repos in Argentina.

[18] The collateral can be either treasury or central bank securities and the haircut is calculated daily by MAE based on their volatility and liquidity, usually ranging from 10 percent to 30%. It is usually around 10 percent for government securities.

the market, affecting therefore the interest rates and monetary conditions of the economy.[19]

Hence, in Argentina, for analytical purposes, the REPO market can be divided in different "segments", according to the participating agents and the type of asset involved as collateral, including transactions: (i) settled by banks and the CB; (ii) between banks (excluding the CB) secured with Treasury bonds and, (iii) between banks (excluding the CB) secured with central bank securities.

Figure 1 depicts the interbank interest rates and BCRA´s policy interest rates during the period under consideration. The CB REPO and reverse REPO interest rates define an "interest rate corridor". In general, both CALL and REPO interest rates are located within the corridor limits. However, for the analyzed period and during several episodes the interbank interest rates crossed that rate corridor limits. It can also be noted that REPO secured market interest rate is usually below the CALL unsecured interest rate.

**Figure 1. Interbank interest rates and Central Bank`s interest rates corridors -%-**

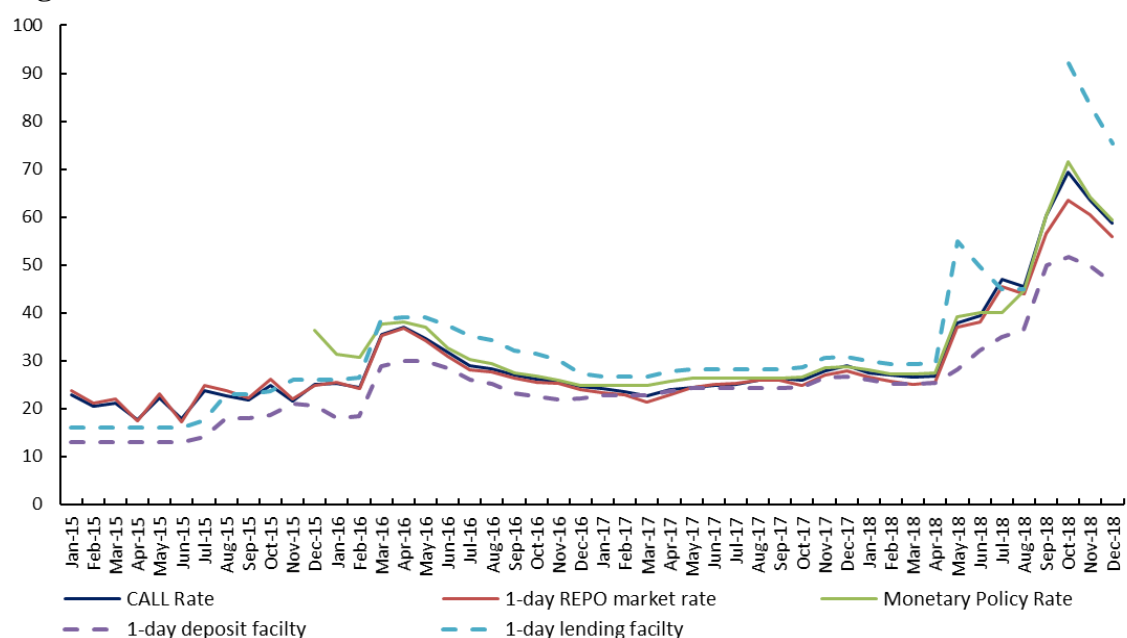

Source: BCRA.

Figure 2 shows the relative importance of each interbank segment. As can be seen, the market volume is mostly explained by the REPOs between banks and the CB. In this segment, most of the transactions are explained by "reverse REPOs", a deposit facility used by the monetary authority to sterilize excess liquidity from the interbank market. In general, banks are not normally willing to ask for lending facilities from the Central Bank as these transactions are considered a bad reputational sign for the market. [20]

---

[19] Since 2002 (in a context of a public debt default), the monetary authority started to issue its own short- and middle-term securities, called LEBAC and NOBAC. These securities were used until 2018, when they were replaced by other Central Bank debt instrument called LELIQ, with a 7-day maturity and that can only be transacted by banks.

[20] In fact, in the period under analysis some of these operations were registered, but they were negligible: only 0,9% of total reverse REPO operations in the 2015-2018 period.

On the other hand, Figure 3 indicates the relative importance of the CALL vis-à-vis the banks REPO market (excluding the CB) during the period under analysis. It can be noted that REPO market was more important than the traditional CALL market during most of the period.

**Figure 2. Interbank Markets Volume** - Monthly average of daily values in Billions of $, constant purchasing power of 2018 -

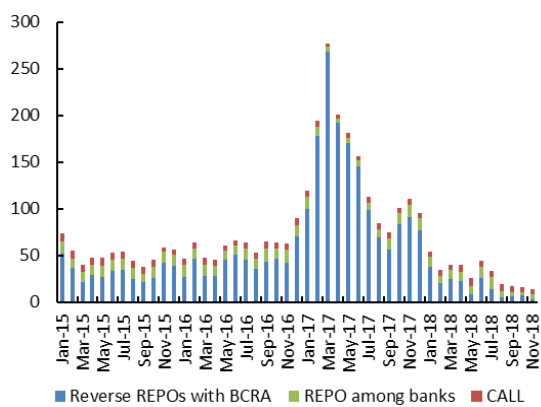

**Figure 3. CALL and REPO Markets Volume** - Monthly average of daily values in Billions of $, constant purchasing power of 2018-

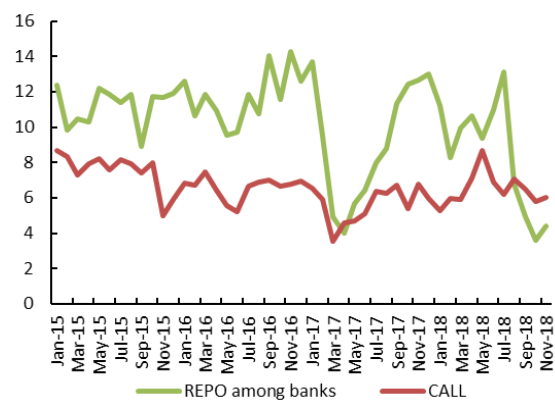

Source: BCRA.

The database includes 78,168 unsecured (CALL) and 150,296 secured (REPO) daily transactions from January 1st, 2015, and December 31st, 2018. Approximately 92% of the transactions in the CALL market and 98% in the REPO market were overnight. In the case of the unsecured market, we use information from the *Siscen* CB database that includes operations among banks on a net daily basis, lender and borrow id (anonymized), volume, maturity, interest rate and currency. On the other hand, the secured market database includes daily information for each transaction pair in the MAE market, including time of the transaction (hour, minute and second), maturity and specific collateral (bond or security), volume and implicit interest rate. In order to analyze the interest rate determinants in both markets controlling for the relevant variables we use a bank balance sheet database, a money market database (with interest rates and regulatory requirements), the current account balance at the CB and the minimum liquidity requirement of each bank.

As can be seen in Table 1 below, a remarkable feature of the time period under analysis is that it includes significantly different monetary policy regimes. In 2015, the first year of the sample, interest rate and capital account controls prevailed.[21] However, both types of controls were liberalized for the rest of the period. The government implemented an inflation targeting regime using the interest rate as the main monetary policy instrument from 2016 onwards. By the end of the sample, in October 2018, the BCRA implemented a monetary base control monetary policy. The primary issuance of BCRA's securities[22] and their interest rate, together with the Pases (active and passive) interest rate corridor represented the monetary policy (or reference) rates between 2016 and 2018.

---

[21] See Forte (2020, p.4).
[22] The LEBAC or CB securities were used for monetary policy implementation during almost the entire period. In mid-August 2018, the BCRA initiated a LEBAC redemption and cancellation program. These instruments were replaced by "LELIQ" (see footnote 19). The redemption process was completed in December 2018. On August 7, the Monetary Policy Committee defined the LELIQ 7-day rate as the monetary policy rate.

**Table 1. Main monetary and macroeconomic events on the period under analysis.**

| Period | Main Events |
| --- | --- |
| 1 January 2015 - 16 December 2015 | Interest rate controls, FX restrictions |
| 9 August 2015 | Primary presidential elections |
| 25 October 2015 | Presidential elections (1st round) |
| 22 November 2015 | Presidential elections (2nd round/Runoff) |
| 10 December 2015 | New Government takes office |
| 17 December 2015 - September 2018 | Inflation Targeting regime progressive implementation |
| 17 December 2015 - 31 December 2016 | Removal of capital controls and interest rate liberalization<br>Monetary Policy Rate: 35-day LEBAC (short-term CB securities) |
| 1 January 2017 - September 2018 | Monetary Policy Rate: 7-day REPO interest rate corridor (mid-point between deposit and lending facilities) |
| August - October 2017 | Midterm elections (primary and general elections) |
| 28 December 2017 | Change in Inflation Target: triggers volatility in the exchange rate market |
| April 2018 - September 2018 | Currency crisis: MPR increased 37.75 p.p. and the peso depreciated 50% |
| September 2018 - December 2018 | Monetary policy regime change: Monetary aggregates control |
|  | Main monetary policy rate: 7-day LELIQ (short-term CB securities) |

As aforementioned, both CB securities and Treasury bonds can be used as collateral assets in REPO market operations. The predominance of Central Bank's securities as the main monetary policy instrument is not the standard international practice, but it is more usual in emerging markets with underdeveloped local debt markets.[23] In the Argentine case, these securities represent the most important banks' tool to manage its liquidity levels, while Treasury bonds are traded in the interbank markets with other objectives (for example, arbitrage transactions, or to exploit carry trade opportunities).

As Figure 4 shows, between 2015 and 2018, 38% of the REPO market operations had Treasury bonds as collateral, while the remaining 62% used CB securities. The CB was involved in 14.5% of the transactions, nearly always using its own securities as guarantee. If these specific operations with the CB are not considered, actually 45% of the transactions carried out between commercial banks were collateralized by Treasury bonds.

---

[23] Rule (2011).

**Figure 4. Collateral assets used in the REPO market operations**
*- Quantity of operations in which each type of collateral was used, as % of total number of operations during the year-*

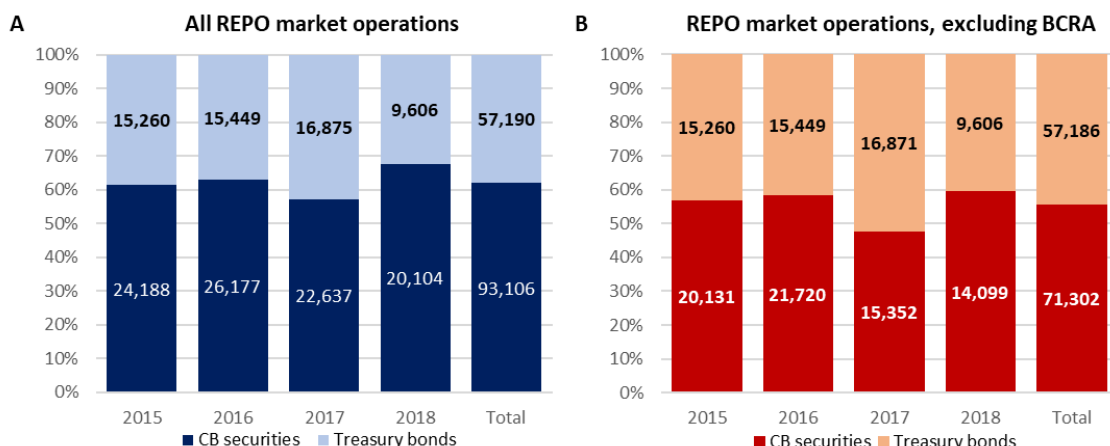

Fig 4. (LHS) Distribution of assets used as collateral in the REPO market. (RHS) Distribution of assets used as collateral in the REPO market, without considering the operations in which the CB was involved.

In practice, this could mean that nearly half of the REPO market transactions between banks (the ones collateralized by Treasury bonds) are settled with intrinsically different purposes than the other half (collateralized by CB securities, which are usually meant for liquidity management). In fact, the difference within the REPO market transactions also arises in terms of the bilateral rates settled in each type of operation, as shown by Figure 5. The bilateral rates of the operations collateralized by Treasury bonds tend to be, on average, lower and more fat-tailed distributed than the transactions backed with CB securities. Additionally, in the latter case the interest rates are more concentrated around the mean interest rate.[24]

**Figure 5. REPO market: Interest rate differentials, according to the collateral asset**

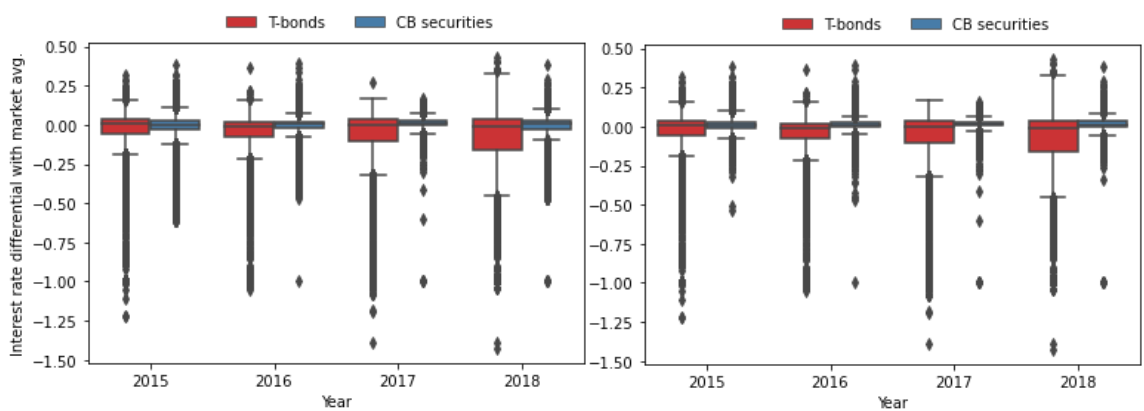

Fig. 5. (LHS) Distribution of the differences between the bilateral rates and the market average rate of the same day in the REPO market. (RHS) Distribution of the interest rate differentials, excluding the transactions in which the CB was involved. (0=T-bonds, 1=CB securities).

---

[24] One of the many reasons behind these differences is that in the REPO market sometimes so-called "special" operations are settled, which essentially are "securities-driven" REPOs. In those cases, the agents are interested in the specific collateral asset, so they are willing to provide cheaper liquidity to obtain it. As the Central Bank consistently intervenes with its own securities as collateral, the "exceptionally far-from-the-mean" bilateral interest rates usually appear when the collateralized assets are Treasury Bonds.

The differences in liquidity conditions of the REPO transactions are statistically significant, depending on whether they have Treasury bonds or CB securities as collateral. To test the difference, we consider a fixed-effects panel data regression model:

$$(2) \qquad r_{ijt} = \beta\, type_{ijt} + \gamma\, X_{ijt} + \mu_i + \delta_j + \theta_t + \varepsilon_{ijt},$$

where **r** is the interest rate of the overnight REPO transaction, **type** is a dummy variable that takes the value 0 if the collateral is a Treasure bond while it is 1 if it is a CB security, **X** is a set of control variables used in Elosegui and Montes-Rojas (2020) to control for liquidity requirements of lender and borrowers (we use the liquidity index defined as in Afonso and Lagos (2015) which is a measure of the liquidity necessity of each bank), amount (logarithm of the volume of the transaction) and maturity (separate dummy variables for one, two, three or four days for the transaction).[25] The model has lender ($i$), borrower ($j$) and day-specific ($t$) fixed-effects and is estimated for each month separately for the period January 2015 to December 2018. This is thus a time-varying regression model for which we are interested in the coefficient of the type of transaction. We also use a different specification where **r** is replaced by **ln(r)**. The results are summarized in Fig. 6 for **r** (left panel) and ln(**r**) (right panel).

**Figure 6. Regression coefficient for interest rate differentials depending on the collateral**

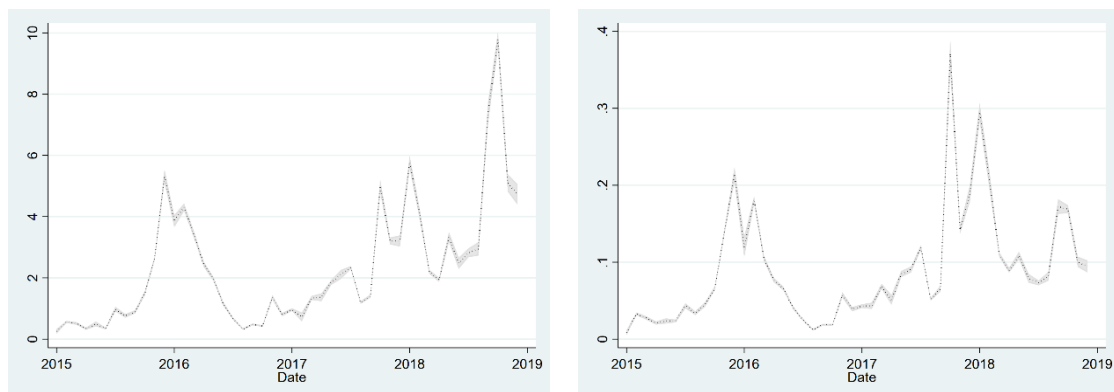

Fig. 6. Regression coefficient of the effect of a dummy variable (0=T-bonds, 1=CB securities) together with 95% confidence interval (robust standard errors clustered by day). (Left) Dependent variable is interest rate. (Right) Dependent variable is log-rate.

Initially assuming no difference by type of collateral in 2015, the econometric results show that macroeconomic uncertainty and episodes of monetary policy tightening are associated with a positive premium of CB securities over Treasury bonds. For example, the 2015-2016 change of government at the national level (the official date is December 10th 2015) is associated with an average 5 percentage points premium or 15% higher interest rate. This spike reduces in the following months. The 2017 midterm elections and the government's announcements of changes in the Inflation targets during December 2017 are also reflected in a higher premium. In 2018, Argentina suffered a severe currency crisis (the Argentinian peso lost 50% of its value that year) triggered by a sudden stop of capital inflows. Coincidentally, the larger premium change is observed in August 2018 (10 percentage points, 19% difference), precisely the month in which the exchange rate tensions peaked. Overall, the results show a positive premium for the entire sample, with larger differences in times of political uncertainty and macroeconomic volatility. A plausible interpretation of these results is that the transactions collateralized by CB securities are driven by intrinsically different purposes (e.g., liquidity management) than the operations secured with Treasury bonds (e.g., arbitrage opportunities).

---

[25] In the case of Gabriele and Labonne (2018a), the authors also regress an interest rate *spread* equation, considering the difference between the annualized (amount-weighted) average interest rate paid in interbank transactions and the deposit facility rate. They found that a larger exposure to higher sovereign risk was associated with higher spreads in their analysis.

In sum, the results indicate that changes in monetary conditions have a significantly different impact on the segment of CB-backed REPO transactions comparing with the impact on the Treasury bonds-secured transactions. This evidence points to the fact that the REPO market is fragmented according to the type of collateral involved in the transactions. In the next sections, we further analyze this divergence between the market segments, considering the underlying network distributions. As we will observe, the fragmentation of the markets is also reflected in significantly different underlying network distributions.

## 4. Empirical results

The interbank market can be represented as a directed network, where the nodes are the banks (including the BCRA, that only operates in the REPO market), while the links are constituted by the transactions among them, in each of the markets. Following the usual practice, the direction of the flows is taken into consideration (therefore, an edge is incoming to the borrower and outgoing from the lender).

The network structures seem to exhibit differences not only between both markets (REPO vs CALL), but also *within* the REPO market (Fig. 7). In the latter case, the structure of the interrelationships apparently differs depending on the participation of the BCRA and according to the collateral asset involved in the transactions (CB securities or Treasury bonds).

# Figure 7. Argentinean interbank markets: Network representations

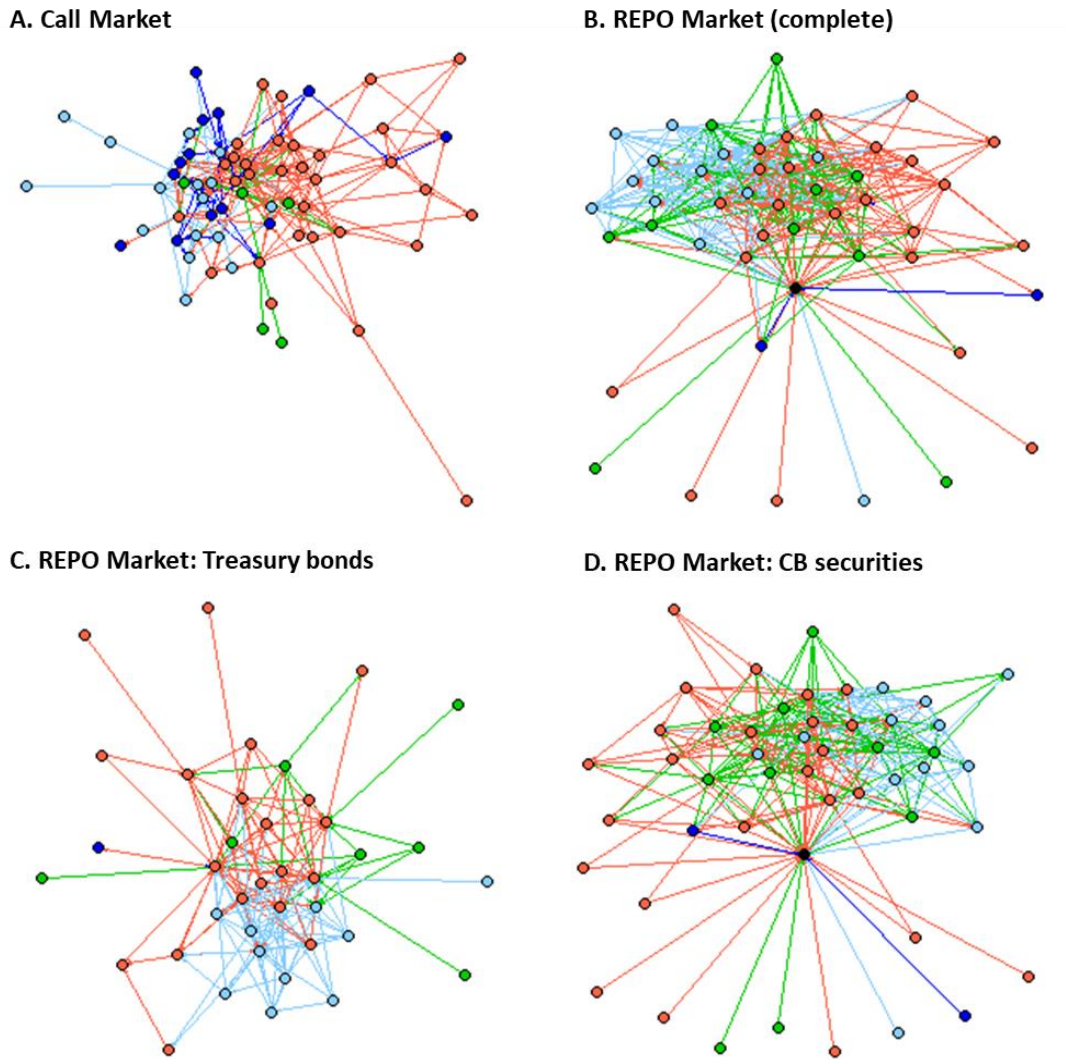

Fig 7. Each node represents a financial entity (green: State-Owned Banks; red: Domestic Private Banks; light blue: Subsidiaries of Foreign Banks; dark blue: Non-Bank Financial Institutions; black: Central Bank). Each edge denotes the existence of at least one loan settled between a pair of entities during the month, and its color is defined by the lender entity. The monthly networks of June 2017 are depicted, as the number of nodes and edges in that month were the most similar to the average of the period under analysis (CALL market: N=65, M=233; REPO market: N=49, M=380). Figures 7C (Lower LHS) and 7D (Lower RHS) display the REPO market network, but only considering in each case the operations collateralized by Treasury bonds or CB securities, respectively. The visualization layout was computed by using the Fruchterman-Reingold algorithm.

In the CALL market network, 64±5 entities (N) established an average of 232±49 links (M) on a monthly basis, while in the REPO market only 49±3 participants were involved, with 398±70 edges (Figure 8). The unsecured market always maintained a higher number of participants than the secured market because of more restrictive barriers to entry in the latter. N remained stable in both markets between 2015 and 2017 but significantly fell since the beginning of 2018 (Figure 8-A).

The number of participants in the REPO market was on average 23% lower than in the CALL market, but nearly 72% more monthly links were created in the former. Hence, the connectivity levels in the REPO networks are significantly higher: the average degree of the nodes ($\hat{k}$) in the REPO market nearly doubles the average degree in the call market, while the density[26] levels in the former are three times higher than in the latter (Figure 8-B and Figure 8-C). Thus, although the unsecured market has more participants, it is not as well connected as the secured market.[27] These conclusions do not substantially change when the operations with the Central Bank are not considered. One plausible interpretation of this result is that the fact that REPO market transactions are secured encourages the establishment of connections between agents which do not necessarily know and trust each other. Instead, the unsecured market demands the creation of "trust" between the agents (after rigorous counterparty risk analysis) before establishing formal links. In addition, the blind electronic platform through which REPO market operations are conducted facilitates transactions among all the participants, while in the case of the CALL market this type of marketplace is absent.

Also, as can be seen in Figure 8-D, the clustering coefficient averaged 51% in the REPO market, substantially above that of the CALL market (16%). In fact, it always remained larger than the clustering levels which would emerge in comparable random networks of the same size[28]. The CALL market's clustering coefficient was not only markedly lower but also fell sharply in 2018, even below the levels expected for a random network of the same size, simultaneously with the crisis that Argentina experienced that year. This points to the conclusion that the lattice of interrelationships in the unsecured market is less stable than in the REPO market, particularly during hard times. Hence, the former market shows signs of being less resilient to negative shocks than the latter.

---

[26] The "density" ($\delta$) of a network quantifies the percentage of the potential links that actually exist, given the number of nodes of the graph: $\delta = \frac{\Sigma_{ij} a_{ij}}{N(N-1)}$ (where $a_{ij}$ is 1 if there is a link between the bank $i$ and the bank $j$ in the given network, and 0 otherwise).
[27] These results are in line, for example, with those obtained by Schumacher (2017) for the Swiss interbank markets.
[28] The average clustering coefficient of a random network with N nodes and $\hat{k}$ average degree is equal to: $\hat{k}$/N (Albert & Barabási, 2002).

**Figure 8. Monthly topological indicators of the REPO and CALL market networks**

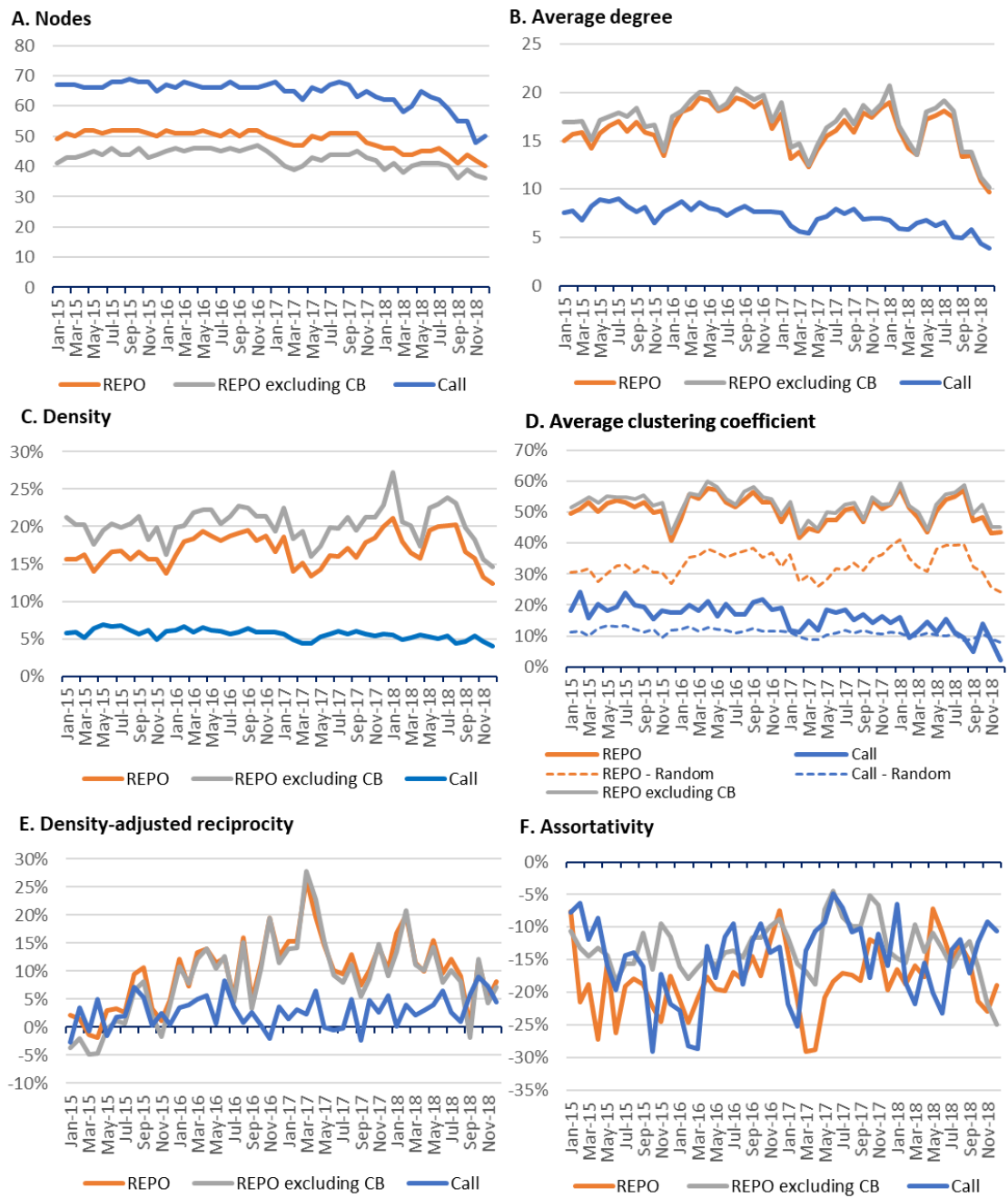

Fig. 8. (A) Number of participants in the network. (B) Average total degree of the nodes. (C) Density. (D) Average clustering coefficient of the nodes in the network, and the expected clustering levels which would emerge from random graphs with the same number of nodes (N) and with the same average degree $\hat{k}$ (the average clustering coefficient of a random network is equal to: $\hat{k}/N$). (E) Reciprocity levels adjusted by the density of the network: (Reciprocity - density) / (1-density), if it is above 0: more reciprocal than a random network, below 0: anti-reciprocal. (F) Assortativity coefficient, computed by using the Pearson correlation coefficient between the degrees of nodes that share links.

Both networks are, on average, more reciprocal than comparable random graphs, which means that banks tend to establish two-way relationships for other reasons than mere randomness (Figure 8 - E). However, the secured market shows higher reciprocity levels than the unsecured market, another indicator of the more embedded relationships in the former.

As it is usual in real-world financial networks (Forte, 2020; Hüser, 2015), both markets are prominently disassortative throughout the period (Figure 8 - F), which means that low-connected banks are more likely to interact with high-degree banks than with other low-degree ones, and vice versa.

Nevertheless, the network structure of the REPO market can be understood as the result of two different underlying networks within that market: one derived from the interactions collateralized by CB securities and other which emerge from the transactions collateralized by Treasury bonds. From Figures 7 - C and 7 - D it is clear that both segments or "sub-markets" present structural differences that deserve to be examined.

First, as mentioned in the introduction, the CB only operates in the REPO market and using its own securities. The monetary authority is the main agent which defines liquidity levels in the market. Consistently, less entities participate in the REPO market using Treasury bonds (37±3) than using CB securities (48±4) (Figure 9 - A.). The average degree of the nodes is higher in the network based on CB securities, especially during 2016 (Figure 9 - B). The connectivity levels in both segments, measured by the average density and clustering coefficient, remained similar throughout the period (Figure 9 – C and 9 - D). However, the network based on Treasury bonds showed more stable metrics, while the market based on CB securities suffered several episodes of volatility: for example, Sep'15-Jan'16 (in coincidence with the change of government ruling party), Nov'16-Jun'17 (a period with a change in the monetary policy framework, as the main reference instrument started to be REPOs instead of LEBACs) , and 2018 (a period characterized by the exchange rate volatility and a sharp reversal of foreign financial flows). Clustering levels remained always above those of a random graph of the same size. In both cases the networks were disassortative, with comparable indicators in this regard (Figure 9 - F).

The most remarkable difference in this topological metrics is related to the Reciprocity: the network based on the transactions collateralized by Treasury bonds is highly reciprocal, with banks establishing numerous two-way links, while, if the transactions collateralized only by CB securities are considered, the network turns out to be nearly neutral or even anti-reciprocal (Figure 9 – E.). This fact is derived from the active presence of the BCRA and the related impact in terms of the network structure of that segment of the market.

**Figure 9. Monthly topological indicators of the REPO market, distinguishing transactions depending on its collateral: CB securities or Treasury bonds.**

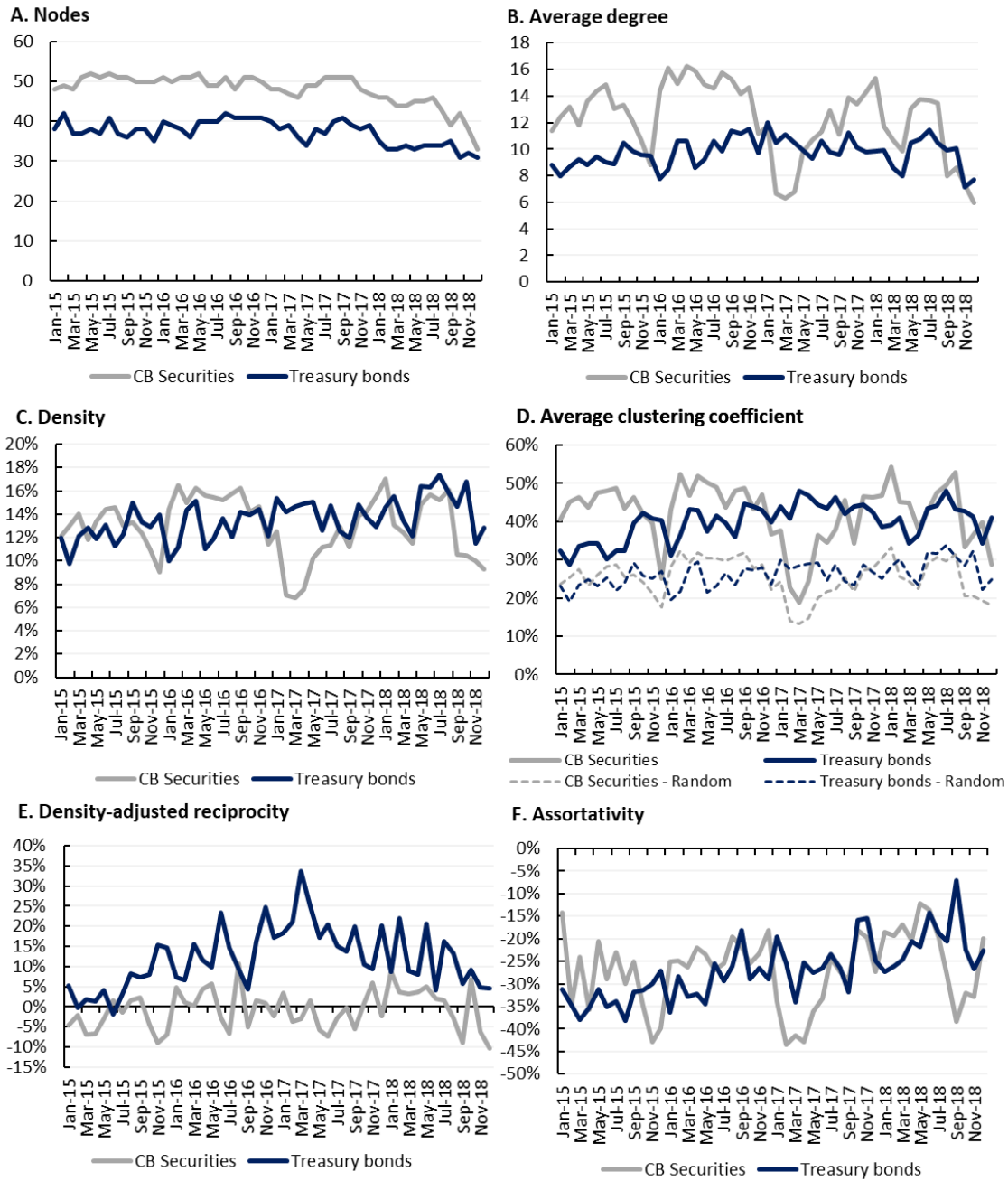

Fig. 9. (A) Number of participants in each sub-network. (B) Average total degree of the nodes. (C) Density. (D) Average clustering coefficient of the nodes in the network, and the expected clustering levels which would emerge from comparable random graphs. (E) Reciprocity levels adjusted by the density of the network (above 0: more reciprocal than a random network, below 0: anti-reciprocal). (F) Assortativity coefficient.

As can be seen in Figure 10, a preliminary analysis of the network degree distribution reflects that they are fat-tailed. After applying the tests developed by Clauset et al. (2009), it can be concluded that the lognormal distribution tends to be the one that best fits the empirical degree distributions of both the CALL and REPO market, outperforming the Poisson and Power Law distributions in the majority of the monthly networks (Table 1). The Lognormal fit is not rejected in 89.6% of the cases in the CALL market, while in the REPO market this hypothesis is not rejected in 91.7% of the cases. This implies that the degree distributions derived from both markets can be characterized as heavy-tailed. That means that they are composed by a few highly connected banks interacting with a myriad of less-connected entities, making both structures vulnerable to targeted attacks or failures of the main agents in the network. This result stresses the relevance of detecting and supervising the central nodes in order to improve the resilience of both networks.

**Figure 10. Complementary Cumulative Distribution Functions of the nodes' degrees**

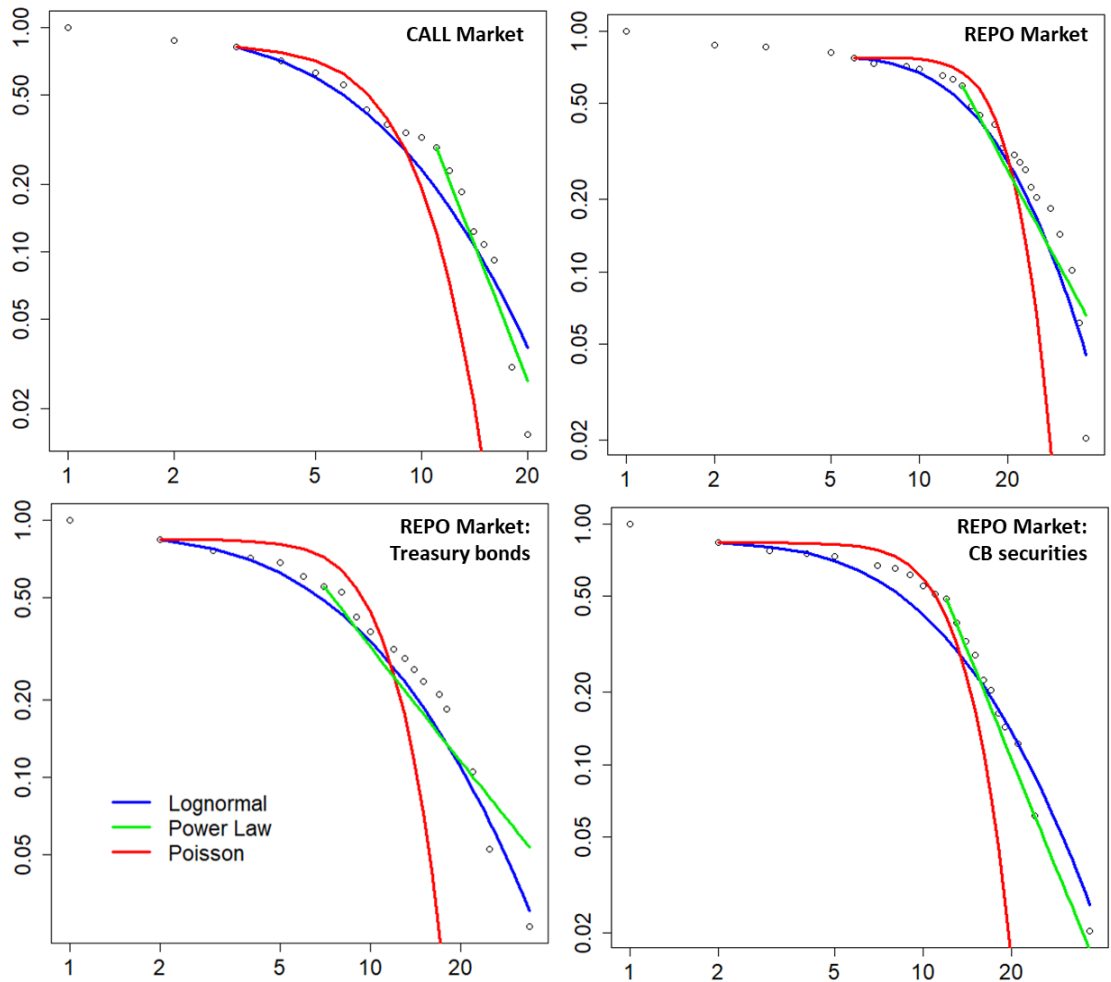

Fig. 10. Networks of June 2017, which were selected for the same reasons detailed in Fig. 7. The best-fit stylized distribution functions are depicted in each chart. Axis are in log-scale.

When the transactions with the BCRA are not taken into consideration, the Poisson hypothesis is rejected in a higher number of months. This evidence points to the fact that the degree distributions of the networks *without* the BCRA are less similar to random graphs and more like fat-tailed than in the case when all the market operations are considered. In fact, when the CB is not an active participant in the networks, those networks become more dependent of a few highly connected nodes. Therefore, as it would be expected, the segment that may be more vulnerable to the failure of its main participants is the one in which the CB is not involved.

**Table 2. CALL and REPO markets: Percentage of monthly networks with a degree distribution that does not reject each null hypothesis**

|  | Call | | | REPO total | | | REPO without BCRA | | |
|---|---|---|---|---|---|---|---|---|---|
|  | Lognormal p>10% | Power Law p>10% | Poisson p>10% | Lognormal p>10% | Power Law p>10% | Poisson p>10% | Lognormal p>10% | Power Law p>10% | Poisson p>10% |
| 2015 | 91.7% | 91.7% | 66.7% | 91.7% | 75.0% | 75.0% | 83.3% | 91.7% | 50.0% |
| 2016 | 83.3% | 66.7% | 83.3% | 100.0% | 100.0% | 75.0% | 100.0% | 91.7% | 50.0% |
| 2017 | 91.7% | 75.0% | 66.7% | 83.3% | 66.7% | 75.0% | 100.0% | 91.7% | 50.0% |
| 2018 | 91.7% | 83.3% | 41.7% | 91.7% | 66.7% | 91.7% | 100.0% | 75.0% | 75.0% |
| Total | 89.6% | 79.2% | 64.6% | 91.7% | 77.1% | 79.2% | 95.8% | 87.5% | 56.3% |
| Avg. Xmin | 4.4 | 9.1 | 10.4 | 11.5 | 21.2 | 24.0 | 12.4 | 20.2 | 19.9 |

Some conclusions from Table 2 are reinforced by the results shown in Table 3. The network based on transactions collateralized by Treasury bonds within the REPO market tends to reject de Poisson distribution as the best fit to its degree distribution in more cases than the network based on CB collateral. The participation of the CB in the network, either directly or indirectly (through its liabilities used as collateral by entities) makes the network structure less similar to a pure fat-tailed distribution, smoothing the *robust-yet-fragile* aspect of that type of networks and hence limiting that type of systemic risk to which the market is subject.

**Table 3. Repo monthly networks, by collateral asset: Percentage of monthly networks with a degree distribution that does not reject each null hypothesis**

|  | REPO - collateral CB | | | REPO - collateral Treasury | | |
|---|---|---|---|---|---|---|
|  | Lognormal p>10% | Power Law p>10% | Poisson p>10% | Lognormal p>10% | Power Law p>10% | Poisson p>10% |
| 2015 | 91.7% | 91.7% | 91.7% | 100.0% | 100.0% | 50.0% |
| 2016 | 83.3% | 100.0% | 41.7% | 91.7% | 75.0% | 41.7% |
| 2017 | 100.0% | 100.0% | 50.0% | 100.0% | 83.3% | 75.0% |
| 2018 | 91.7% | 83.3% | 100.0% | 83.3% | 91.7% | 75.0% |
| Total | 91.7% | 93.8% | 70.8% | 93.8% | 87.5% | 60.4% |
| Avg. Xmin | 11.6 | 17.4 | 15.1 | 4.3 | 11.5 | 12.4 |

The results indicate that in the network of REPO operations collateralized by Treasury bonds, the key participation of central agents is relatively more important to the well-functioning of the market than in the case of the transactions backed by CB securities.

Finally, in order to confirm these findings, we introduce an additional procedure, to address the issue of detecting the best fit for each monthly degree distribution. The procedure can be summarized in three steps:

(i) Apply the method described by Clauset et al. (2009) to the empirical degree distributions, as it was done in Tables 1 and 2.

(ii) If more than one theoretical distribution is not rejected, a classical Vuong Test is performed to define which the better fit is.
(iii) If Vuong Test does not provide enough evidence to select one or other distribution, then the log-likelihood associated to each fit is compared to decide the better one.

The results of this procedure for each (monthly) network of both markets are summarized in Figure 11. The evidence reinforces the outperformance of the lognormal fit over the others. Considering the total period under analysis, the Call market and the REPO market network structure does not differ significantly in this regard: in nearly 70% of the months the best fit is achieved by the Lognormal distribution, while in the other 30% the Poisson distribution is better.

**Figure 11. Interbank (monthly) networks best described by each parametric distribution function (in % of total months).**

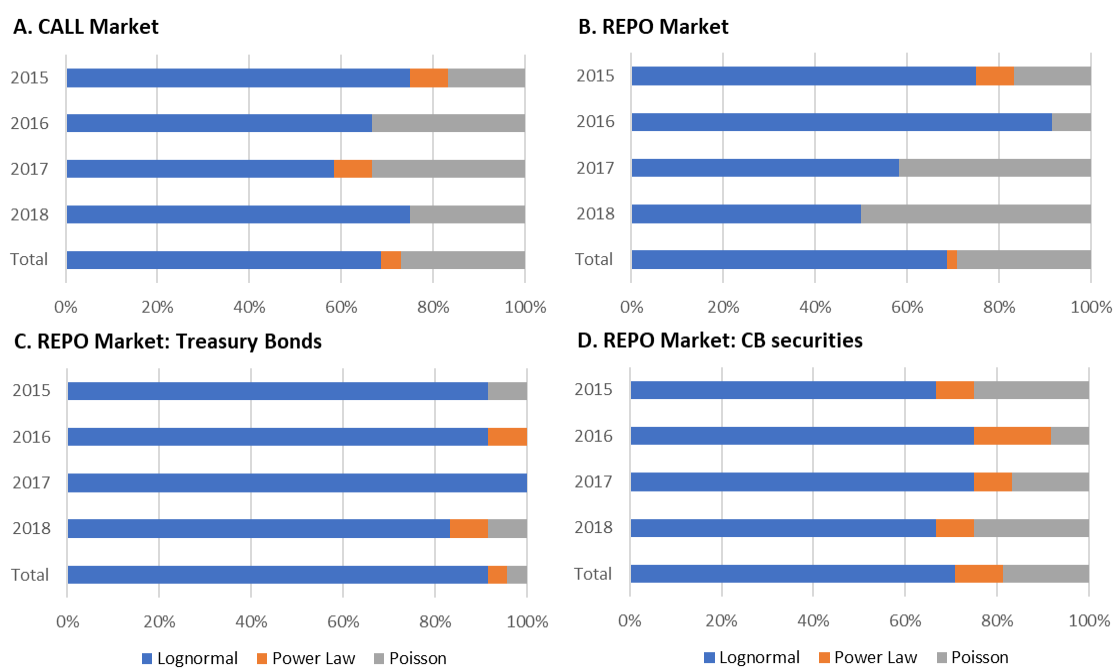

It should be noted that the REPO market network shows signs of having experienced a structural change during 2018, as its topology seems to have partially "randomized" in an economic period characterized by a Balance of Payment crisis and several exchange rate depreciations episodes. Conversely, during the same period the CALL market rejects most of the time the Poisson hypothesis. This reflects that both segments reacted in different ways under similarly stressful events.[29]

Figures 11 C and D, indicate that the network based on the REPO market transactions collateralized by Treasury bonds is significantly better described by fat-tailed distributions than the network based on CB-securities-backed transactions. In fact, for the latter, in some more cases the random network performs better. The difference has

---
[29] The observed changes under a stress event may deserve better attention from the financial stability perspective that is beyond the present analysis.

implications in terms of the fragility of these segments of the interbank market. The former is more vulnerable to the failure of its main agents, while the market based on CB securities seems to be slightly more resilient to these types of events. In the latter, the participation of the BCRA (directly or indirectly, through its securities) proves to be important in terms of stability.

5. **Concluding remarks and discussion**

This paper examines the secured (REPO) and unsecured (CALL) interbank markets of Argentina applying a complex networks approach to analyze market fragmentation. The empirical analysis is performed using a unique database spanning from 2015 to 2018, a particularly volatile period for the Argentinean economy.

Based on standard topological metrics (like the average degree, density and clustering coefficients), it is found that, although the secured market has less participants, its nodes are more densely connected than in the unsecured market. In addition, the interrelationships in the latter are less stable, as it was witnessed during the 2018 currency crisis, making its structure more volatile and vulnerable to negative shocks.

In general, the main topological indicators of the Argentinean interbank markets are in line with those found in other Latin American countries: disassortative behavior, relatively short average distances, low density levels, similar clustering coefficients and heavy-tailed degree distributions.

The analysis identifies two "hidden" underlying sub-networks within the REPO market: one based on the transactions collateralized by Treasury bonds (REPO-T) and other based on the operations collateralized by CB securities (REPO-CB). The connectivity indicators were significantly more stable in the REPO-T market than in the REPO-CB segment, as the latter is evidently more correlated with the liquidity swings defined by the Central Bank. The changes in monetary policy stance and monetary conditions seem to have a substantially smaller impact in former than in the latter "sub-market". Hence, the connectivity levels within the REPO-T market remain relatively unaffected by the (in some period pronounced) swings in the other segment of the market.

The reciprocal relationships in the REPO-T segment are significantly more frequent than in the REPO-CB, showing that BCRA's "one-way operations" in the REPO market crucially shapes the type of relationships established in the latter sub-network. Meanwhile, the fact that relationships are highly reciprocal in the subset of transactions collateralized by Treasury bonds reflects that this sub-network is significantly less affected by the participation of the BCRA in the REPO market.

In terms of financial stability, the distribution function that best fits the empirical degree distributions in both the secured and the unsecured market is the lognormal, a fat-tailed distribution. As a result, the networks are composed by a few highly connected banks jointly with multiple entities with a significantly lower degree. Given this network structure, the highly interconnected banks are key for the stability of the markets, hence the regulation and supervision should focus on the well-functioning of these central agents to preserve the system's stability. However, the participation of the Central Bank in the REPO market subdues somewhat this conclusion. When the transactions with the BCRA are not considered, the REPO market degree distribution becomes closer to a fat-tailed distribution. But when the transactions with the BCRA are considered, this conclusion is less categorical, and the Poisson hypothesis gains some support. This evidence implies that the participation of the Central Bank in the REPO market alters the structure and the underlying risks of the network. In this sense, the markets in which the Central Bank does not intervene directly (the CALL market and REPO-T) would deserve a different treatment from the point of view of financial stability supervisors. Nevertheless, it is important to note that the direction of the "causality" between the markets' risk structure and the Central Bank active participation (or not) is not straightforward. An extensive presence of the monetary authority in these markets could also undermine the development of a denser lattice of interrelationships among private entities, which could, in turn, mitigate the risks of a fat-tailed degree distribution in the networks.

Overall, these differences seem to be reflecting that the transactions collateralized with CB securities may have different motivation than those collateralized by T-bonds. In fact, the REPO-CB market appears to be more related to liquidity management activities, while the REPO-T market could be mainly motivated by arbitrage operations or the demand of certain bonds for specific reasons, not necessarily related to liquidity management decisions.

The existence of two sub-markets with different structure makes it difficult to fully interpret or extrapolate the implications of the average interest rate that emerges from the REPO market. In fact, two very different types of decisions are condensed in that same price. In contrast, the CALL unsecured market is more homogeneous, with an interest rate reflecting a "clearer" reference of the "cost of money" in the financial system. This "fragmentation" of the REPO market has multiple implications. For example, the dichotomy posed by this "fragmented market" behind the formation of the REPO market interest rate jeopardizes the development of an "interest rate forward market" for this rate (or broader derivative markets of this rate in general), a pending task in Argentina. This issue could be solved if the Central Bank computes (and publishes) an interest rate trying to capture the average rate of the operations that are settled only for liquidity management purposes, which should be "near" to the monetary policy rate, defined by a certain (empirically defined) threshold. Such indicator could be used as a reference to settle contracts, avoiding the biases introduced by, for example, special REPO market operations and/or arbitrage transactions derived from extraordinary conditions or attributes of specific collateral assets. These issues constitute an agenda for further research.